# Correlation of magneto-electric coupling with crystal structure in Dy doped hexagonal $YMnO_3$

A. K. Singh and S. Patnaik

School of Physical Sciences,

Jawaharlal Nehru University, New Delhi – 110067, India

S. D. Kaushik and V. Siruguri

UGC-DAE-Consortium for Scientific Research Mumbai Centre,

R5 Shed, Bhabha Atomic Research Centre, Mumbai – 400085, India

**Abstract**

Coexistence of mutually influencing ferroelectric and magnetic ordering at low temperature in hexagonal phase of rare-earth manganites gives rise to interesting physical phenomena. Here, we report a detailed analysis of powder neutron diffraction (ND) studies on polycrystalline $Y_{1-x}Dy_xMnO_3$ (x = 0, 0.05) as a function of temperature. Our analysis shows that Dy doping at Y site leads to modification in lattice constants: unit cell shrinks in *c* direction, but expands in *ab* plane, unit cell volume increases and effective Mn magnetic moment also reduces. All the Mn-O bond lengths and O-Mn-O bond angles shows significant variations leading to reduction in tilting of $MnO_5$ polyhedra. Dielectric measurements on these samples show that, due to Dy doping, inverse S-shape anomaly shifts to the lower temperature side and magnetoelectric coupling decreases. Magnetic measurements confirm the decrease in the frustration of Mn triangular lattice. Temperature dependent specific heat measurement shows anomaly at Néel temperature ($T_N$) and results indicate a decrease in entropy due to Dy doping. These results are understood in terms of modification in magnetoelectric coupling triggered due to Dy induced strain in Y plane of $YMnO_3$.

**Introduction**

The correlation between magnetic, electric and elastic properties has been pursued for a long time in condensed matter physics [1-5]. Magneto-electric coupling in multiferroics, which possess electric and magnetic ordering simultaneously, has been extensively explored but with a small rate of success [6-9]. These materials hold the promise for potential technological applications in nonvolatile multiple-state data storage, magnetic field sensors, transducers, and actuators [10-13]. These novel applications have stimulated the search for new multiferroics and highlighted the need for a thorough understanding of physics behind these exciting materials. The non-trivial spin-lattice coupling in multiferroics has been observed through various forms, such as linear and bilinear magneto-electric effects [14], polarization change through field-induced phase transition [15, 16], magneto-dielectric effect [17, 18], and dielectric anomalies at magnetic transition temperatures [19, 20]. The major drawback of currently known multiferroic compounds is that they undergo magnetic and ferroelectric phase transitions at much different temperatures, resulting in a weak coupling between both the orders [21, 22].

Lanthanide-based manganites $RMnO_3$ (R=Tb, Dy,…Lu) have attracted a great deal of attention because of their flexibility in forming orthorhombic or hexagonal crystal structures, depending on the lanthanide ionic radius $r_R^{+3}$. *R*$MnO_3$ stabilizes in an orthorhombic phase (*Pbnm*) [23] when $r_R^{+3} \geq r_{Dy}^{+3}$ and a hexagonal phase (*P6*$_3$*cm*) [24] is formed when $r_R^{+3} < r_{Dy}^{+3}$. The orthorhombic *R*$MnO_3$ (*R*=Tb, Dy, and Gd) are reported to be antiferromagnetic (AF) with $T_N$ ~ 40 K and ferroelectric (FE) phase appears below $T_N$. Ferroelectric polarization in this class of compounds is governed by lattice modulation accompanied by the AF order [25-27]. On the other hand, the hexagonal *R*$MnO_3$ (*R*=Ho, Er, Tm, Yb, Lu, and Y) compounds are of particular interest due to coupling between ferroelectricity and ferromagnetism and their possible control by the application of magnetic and/or electric fields [28]. These compounds order ferroelectrically around 600 K and AF ordering temperature i.e. Néel temperature ($T_N$) ranges from 56 to 120 K [29-31].

Of the hexagonal rare earth manganites, $YMnO_3$, one of the most intensively studied compound, is of particular interest because it has simpler crystalline and magnetic order, which may facilitate the unraveling of the complex physics of multiferroicity. The crystal structure of hexagonal $YMnO_3$ consists of alternating $MnO_5$ triangular bipyramidal layers and rare-earth (Y) layers in the out-of-plane direction [28, 32, 33]. A striking feature of this structure is a 2D in-plane $Mn^{3+}$ (S=2) stacked triangular lattice causing geometrical frustration with frustration parameter $f$ = 5 to 10, where $f = \theta_{CW}/T_N$, with $\theta_{CW}$ and $T_N$ as Curie-Weiss and Néel temperature respectively [19, 28, 33, 34]. Due to these competing AF superexchange interactions among $Mn^{3+}$ spins, each of the three sublattices are oriented at 120° from the other and classical magnetic ground state acquires noncollinear spiral configuration [33, 35]. The coupling of these triangular lattices of Mn layers along the *c*-axis is difficult to resolve from macroscopic measurements but it has been investigated for various hexagonal manganites by ND experiments [36]. Giant magnetoelastic coupling has been observed in the ferroelectric as well as AF phases and modification in this coupling due to external magnetic has also been studied [28, 33].

In hexagonal $RMnO_3$, the $R^{3+}$ site may have non-magnetic (e.g. $Y^{3+}$) [32, 33] or magnetic (e.g. $Tb^{3+}$, $Dy^{3+}$) [37, 38] atom. The magnetic $R^{3+}$ ions order at very low temperatures (below 15 K) and this makes the frustrated Mn triangular sublattice more complicated [39]. From ND studies, it is well known that $R^{3+}$ ions occupy two inequivalent crystallographic positions [2(a) and 4(b) represented as *R*1 and *R*2 respectively] whereas the $Mn^{3+}$ ion occupies the 6(c) position [19, 32, 33]. The magnetic moments on *R*1 and *R*2 sites are considered to be of different strengths and independent of each other [40]. The geometric arrangement of sub-lattice remains the same for different rare earth ions, but the magnetic properties can be quite different depending on the size of $R^{3+}$ ion. Interestingly, the rare-earth moments at the 4(b) sites are usually ordered and directed parallel to the c axis (i.e. perpendicular to basal plane) causing the inter-sublattice coupling to be frustrated and magnetic phase diagram becomes more complex [41].

In order to study the microscopic origin of multiferroicity in $YMnO_3$, various doping studies had been performed at both magnetic Mn site and non-magnetic Y site [42-45]. It is observed that multiferroic property of $YMnO_3$ is governed by both the magnetic Mn and non magnetic Y atoms [46]. It would be interesting to see how the multiferroic property gets modified while transforming from hexagonal to orthorhombic structure. Orthorhombic $DyMnO_3$, a known multiferroic, contains larger Dy, so doping Dy at smaller Y site is expected to produce significant impact on structure and multiferroic property. The partial replacement of Y by Dy results in a change of orbital occupation ($5s^2\ 4d^1$ to $6s^2\ 4f^{10}$) and ionic radius (90 to 91.2 pm). From magnetization measurements of $DyMnO_3$, it was concluded that the $Dy^{3+}$ moments order ferrimagnetically below 7 K [47-49]. Nevertheless, the magnetic structure and the corresponding magnetic symmetry remained unknown for orthorhombic $DyMnO_3$ [35]. Thus we can understand emerging physical properties due to doping of large Dy at Y site in hexagonal $YMnO_3$.

In this paper, we demonstrate that in hexagonal $YMnO_3$, larger $Dy^{3+}$ ion could be doped in place of smaller $Y^{3+}$ ion. Modification in the crystal structure due to 5% Dy doping was studied by X-ray diffraction and temperature dependent ND measurements. It is observed that doping results in significant variation in microscopic parameters such as lattice constants ***a*, *c*,** unit cell volume, ordered Mn magnetic moment, Mn-O bond lengths and O-Mn-O bond angles. Magnetic measurement clearly shows the decrease in the AF transition temperature for x = 0.05 in $Y_{1-x}Dy_xMnO_3$ as compared to $YMnO_3$. These results are further correlated with the results of temperature dependent magneto-dielectric study. Specific heat measurement in presence of magnetic field shows the evidence of ordering of $Dy^{3+}$ moment below 15 K. Due to higher neutron absorption coefficient ($\sigma_a$ ~ 994 barn) of the naturally occurring isotope of Dy, it is challenging to determine the magnetic structure of $Y_{1-x}Dy_xMnO_3$ by ND and hence the Dy doping is limited to only 5% in the present work.

**Experimental techniques**

Polycrystalline samples of $Y_{1-x}Dy_xMnO_3$ (x = 0, 0.05) were synthesized by standard solid-state reaction method at ambient pressure. $Y_2O_3$ (99.99%), $Dy_2O_3$ (99.99%) and $MnO_2$ (99.99%) were thoroughly mixed ground in a ratio of 1:2 to achieve the stoichiometry of $Y_{1-x}Dy_xMnO_3$ (x = 0, 0.05). The mixed powders were compacted and heated to 1100 °C for 13 h and later they were annealed at 1300 °C for 12 h with intermediate grindings in order to ensure the homogeneity and density of pellets for dielectric measurements. Room temperature powder X-ray diffraction (XRD) studies of the samples were performed using a Bruker D8 X-ray diffractometer with Cu $K_\alpha$ radiation ($\lambda$=1.5406 Å). The ND measurements were carried out on powder samples using the multi-position sensitive detector based Focusing Crystal Diffractometer (FCD) of UGC-DAE Consortium for Scientific Research Mumbai Center at Dhruva reactor, Mumbai (India) at a wavelength of 1.48 Å in the temperature range of 10 to 300 K.

The ND patterns were analyzed using Rietveld method and the refinement of both crystal and magnetic structures was carried out using the FULLPROF program [50]. Temperature and frequency dependent dielectric measurements on the pellets were performed using a QUADTECH 1920 precision LCR meter using a cryogen-free low temperature high magnetic field facility. Magnetization and specific heat data were recorded in a Quantum Design

Magnetic Property Measurement System (MPMS) SQUID and Physical Property Measurement System (PPMS), respectively.

**Results and discussion**

Fig 1(a) shows the room temperature XRD pattern for $Y_{1-x}Dy_xMnO_3$, x = 0, 0.05 samples. For both the samples, all peaks are satisfactorily indexed in hexagonal crystal structure with space group *P6₃cm*. Due to doping of larger $Dy^{3+}$ in place of smaller $Y^{3+}$, all the lattice peaks shift to the lower 2θ values which signifies increase in lattice volume. The inset of Fig 1(a) shows the XRD pattern of x = 0.05 sample in the 2θ range of 15-70°.

Fig 1(b) shows Rietveld refinement of room temperature ND patterns of $Y_{1-x}Dy_xMnO_3$ x = 0, 0.05. It is observed that there is no change in the crystal structure from room temperature to 10 K, thus indicating that the hexagonal structure is retained up to 5 % doping of Dy in $YMnO_3$. Lattice constants ***a*** = 6.149 Å and ***c*** = 11.363 Å for $YMnO_3$ are in a good agreement with the data available in literature [42, 45]. Due to Dy doping, the lattice constant ***a*** and unit cell volume increase whereas lattice constant ***c*** decreases. ND pattern of the doped sample does not show the presence of any impurity peaks. Structural parameters after the Rietveld refinement of ND patterns for $Y_{1-x}Dy_xMnO_{3,}$ x = 0, 0.05 at room temperature are shown in Table 1. In the inset of Fig 1(b), ND patterns at 10 K for both the samples are shown and the magnetic Bragg peaks are marked by (*). Low temperature ND patterns were recorded in warming cycle. ND patterns below $T_N$ show the expected magnetic intensities, in particular the peak (101) which is forbidden by the space group $P6_{3cm}$. Appearance of this peak is a signature of the AF below $T_N$ characterized by the propagation vector *k*=0. For the magnetic refinement of the ND pattern at 10 K, the structure described by the basis vectors of the irreducible representation $\Gamma_1$ was used.

In the entire temperature range measured during warming cycle from 10 K to 300 K, lattice constant ***a*** shows increasing trend and ***c*** shows decreasing trend in for both x = 0 and 0.05 samples. Shrinkage in ***c***-axis due to Dy doping is expected to buckle or release the Y plane, and this strain may possibly modify the $MnO_5$ polyhedron tilt. The effect of this reorientation can be observed in ordered state magnetic moment of Mn ions located on z = 0 and z = 1/2 planes. Variation of lattice constants, unit cell volume, and ordered Mn ion magnetic moment as a function of temperature is depicted in Fig 2(a)-(c). We observe significant suppression in Mn ion magnetic moment in $Y_{1-x}Dy_xMnO_3$, x = 0.05 as determined from the magnetic Rietveld refinement. It is clearly observed that for x = 0.05, the ordered Mn ion moment decreases much faster as compared to x = 0 signifying the reduction of $T_N$ for x = 0.05. Dielectric measurements on x = 0.05 confirm decrease in AF transition temperature ($T_N$) to ~ 55 K as well as in magnitude of real part of dielectric constant. In case of $RMnO_3$, it has been reported that due to the tilting of $MnO_5$ polyhedra and buckling of *R* plane, *c*-axis gets elongated [51, 52] but in the present case, ***c*** - axis decreases with respect to Dy doping which suggests tilting and buckling are reduced.

From Fig 3(a)-(d), we see that, on one hand, Mn-O1 and Mn-O3 bond lengths reduces on induction of 5% Dy at Y site at all temperatures below and above the ordering state. While on the other hand, Mn-O2 and Mn-O4 bond lengths increase in the same situation. It is known that in pure $YMnO_3$ structure, Mn atoms are situated at z=0 and z=1/2 plane and each Mn is surrounded by five oxygen atoms forming $MnO_5$ trigonal bipyramidal structure. O1, O2 represent apical oxygen ($O_{ap}$) atoms and O3, O4 represent equatorial oxygen ($O_{eq}$) atoms. Variation in Mn bond length with equatorial oxygen is more prominent in the ordered state. Thus, we clearly observe reorientation in $MnO_5$ polyhedra. Bond angles are depicted in Fig 3(e)-(j). This reorientation results in increase of O1-Mn-O2, O1-Mn-O3, O1-Mn-O4, and O2-Mn-O3 bond angles while O2-Mn-O4 bond angle decreases all over temperature range from 10 K to 100 K . Tilting of $MnO_5$ polyhedra is represented by the angle between O(1)–O(2) (apical oxygen) bond and c-axis, which is reflected in O1-Mn-O2 bond angle, Fig 2 (e) clearly show the increase in O1-mn-O2 bond angle, thus signifying the modification in the magnitude of tilting of $MnO_5$ polyhedra. The reduction in the tilting and buckling in $MnO_5$ polyhedra may be induced due to doping of large $Dy^{3+}$ in place of smaller $Y^{3+}$.

The magnetic susceptibility studies of $Y_{1-x}Dy_xMnO_3$ were done at $\mu_0H$=0.5 T in zero field cooled (ZFC) condition and presented in Fig 4. Inset shows the reciprocal magnetic susceptibility ($1/\chi$) as a function of temperature which, for both the samples, follows Curie-Weiss behavior with the effective paramagnetic moments 4.67 $\mu_B$ and 4.95 $\mu_B$ for x = 0, 0.05, respectively. These values are in a good agreement with the free $Mn^{3+}$ ion value, 4.9 $\mu_B$ (for S=2). The extrapolated paramagnetic temperatures are -330 K for x = 0 and -225 K for x = 0.05 and the degrees of frustration (*f*) are 5.5 and 3.8, respectively. Thus, Dy doping in $YMnO_3$ decreases the frustration factor significantly. At low temperatures, the Dy moments start ordering and due to this reason the moment for Dy doped sample is larger as compared to $YMnO_3$.

Temperature dependent dielectric constant ($\varepsilon$) of the sintered pellets of $Y_{1-x}Dy_xMnO_3$ across the AF transition at a frequency of 1 kHz and at $\mu_0H$ = 0 T is shown in Fig 5(a). For x = 0, a clear signature of a weak first order phase transition (AF ordering) is clearly evident at T ~ 62 K, whereas for x = 0.05, transition is observed at T ~ 58 K. For Dy doped samples, the dielectric constant decreases considerably and also the inverse S-shape anomaly shifts to lower temperature side (shown in the inset of fig 5 (a)). For both the samples, $\varepsilon$ increases with increasing temperature and the increment is sharp above 200 K which persists up to room temperature. At room temperature, $\varepsilon$ for both the samples becomes almost equal, pointing towards less effect of 5% Dy doping at the room temperature. Magnetocapacitance (MC) for both the samples is shown in Fig 4(b) in presence of $\mu_0H$= 0, 3, 5 T. The strength of MC has been determined using the relation, $MC = \frac{\varepsilon(T,H) - \varepsilon(T,H=0)}{\varepsilon(T,H=0)} \times 100$. This strength for x = 0 at 65 K in the presence 3 T magnetic field is 0.33% and for 5 T, it is 0.5%, whereas for x = 0.05 at 58 K, which is $T_N$, the strengths of MC at 3 T and 5 T are 0.05% and 0.16%, respectively. Hence, from field dependent dielectric measurements on both samples, we conclude that the magnetoelectric coupling decreases due to doping of larger ionic radii Dy in place of smaller Y. The change in the dielectric constant with the magnetic field is purely of capacitive origin and the possibility of leakage current and presence of MC due to magnetoresistance has been ruled out [28, 53]. In case of $YMnO_3$, field dependent ND experiments clearly show that magnetic field suppresses the tilting of $MnO_5$ polyhedra and due to this reason, ε decreases in presence of magnetic field. In case of Dy doped $YMnO_3$, doping exerts a chemical pressure on the unit cell and in presence of magnetic field, the change in dielectric constant is even less as compared to $YMnO_3$. Clearly, due to Dy doping, the unit cell becomes more rigid and less prone to external magnetic field. The changes in the dielectric constant due to magnetic field in the paramagnetic state provide the evidence of dominant magnetoelastic coupling over magnetoelectric coupling [28]. As compared to $YMnO_3$ the magnetoelastic coupling decreases significantly in Dy doped samples.

In order to study the specific heat properties of the pure and doped samples, we measured the heat capacity from 2 to 90 K. The raw data, shown in the main panel of Fig 6(a) displays a clear peak indicative of the AF transition ($T_N$). The peak position in the heat capacity data shifts from 62 K for x = 0 to 58 K for x = 0.05. Due to ordering of $Dy^{3+}$ below 15 K, an anomaly in the specific heat is observed and it is expected that for doped samples magnetic entropy will be more. In the inset, we also plotted the magnetic specific heat $\Delta C_P/T$ after subtracting phonon contributions. In order to calculate magnetic entropy from the raw heat capacity data, we have estimated the phonon contribution using the Debye model with two Debye temperatures ($\theta_1$=413±5 K and $\theta_1$=825±6). This two-Debye-temperature model is used because there are two kinds of elements in our samples: three relatively heavy elements (Y, Dy, and Mn) and a relatively light element (O), which are expected to give rise to distinctively different Debye temperatures [44]. The anomaly in the specific heat for Dy doped sample at low temperature becomes more prominent after phonon subtraction. By subtracting the phonon contribution from the raw data, we have obtained the total magnetic entropy of 11.45 J $mol^{-1}$ $K^{-1}$ for $YMnO_3$, 12.78 J $mol^{-1}$ $K^{-1}$ for x = 0.05. We can ascribe the increased magnetic entropy to the doped magnetic $Dy^{3+}$ in place of non magnetic $Y^{3+}$ within the resolutions of our experiments. We have also studied the magnetic field dependent heat capacity as a function of temperature for both the samples. When a magnetic field $\mu_0H$ up to 5 T is applied, no changes are observed in the transition temperatures for both the samples. The specific heat for $YMnO_3$ remains unperturbed due to external magnetic field in the entire temperature range 2 - 90 K. Whereas for x =0.05, a significant variation below 15 K is observed even for 1 T magnetic field, shown in Fig 6(b). These variations become more prominent at even higher magnetic field ($\mu_0H$ = 5 T). It is to be noted that for 1 T magnetic field, the magnetic entropy increases (12.83 J $mol^{-1}$ $K^{-1}$) as compared to zero field and when the field is increased to 5 T, it decreases (12.47 J $mol^{-1}$ $K^{-1}$).

**Conclusion**

In summary, we have investigated the crystal structure of $Y_{1-x}Dy_xMnO_3$ (x = 0, 0.05) by means of XRD and temperature dependent neutron diffraction experiments. ND measurements revealed that the size of the ordered moment is reduced from the expected value of $Mn^{3+}$ and this value further decreased when Dy is doped at Y site. These magnetic properties are caused by strong geometrical frustration of spins on the triangular lattice of Mn ions which is well supported by magnetization measurements. Temperature dependence of microscopic parameters clearly establishes that Dy suppresses the tilting of $MnO_5$ polyhedra and the buckling of Y planes. Field dependent dielectric data reveal that substantial magnetoelectric coupling is present for $YMnO_3$ and it decreases considerably when Dy is doped at Y site. Specific heat measurement also confirms the reduction in entropy.

**Acknowledgement**

We thank the Department of Science of Technology, Government of India, for the financial support under the FIST program to JNU, New Delhi and the UGC-DAE Consortium for Scientific Research, India for providing support through a Collaborative Research Scheme for accessing the neutron scattering facilities at NFNBR, BARC, Mumbai. We thank Dr. V. Ganeshan, UGC-DAE Consortium for Scientific Research, Indore, India for the specific heat measurement and very useful discussions. A.K.S. would like to thank CSIR, India for the fellowship.

## References

1) D. Meier, M. Maringer, T. Lottermoser, P. Becker, L. Bohatý, and M. Fiebig, Phys. Rev. Lett. **102**, 107202 (2009).

2) A. Pimenov, A. M. Shuvaev, A. A. Mukhin and A. Loid, J. Phys.: Condens. Matter **20**, 434209 (2008).

3) C. N. R. Rao and C. R. Serrao, J. Mater. Chem. **17**, 4931 (2007).

4) A. K. Singh, M. Snure, A. Tiwari, and S. Patnaik, J. Appl. Phys. **106**, 014109 (2009).

5) D. Khomskii, Physics **2**, 20 (2009).

6) J. M. Rondinelli, M. Stengel, and N. A. Spaldin, Nat. Nanotech. **3**, 46 (2008).

7) Y. Tokura, J. Mag. & Mag. Mater. **310**, 1145 (2007).

8) T. Kimura, Annu. Rev. Mater. Res. **37**, 387 (2007).

9) M. Stengel, D. Vanderbilt, and N. A. Spaldin, Nat. Mater. **8**, 392 (2009).

10) J. Zhai, Z. Xing, S. Dong, J. Li, and D. Viehland, Appl. Phys. Lett. **88**, 062510 (2006).

11) M. Vopsaroiu, J. Blackburn, A. Muniz-Piniella, and M. G. Cain, J. Appl. Phys. **103**, 07F506 (2008).

12) F. Yang, M. H. Tang, Z. Ye, Y. C. Zhou, X. J. Zheng, J. X. Tang, and J. J. Zhang, J. Appl. Phys. **102**, 044504 (2007).

13) N. Hur, S. Park, P. A. Sharma, J. S. Ahn, S. Guha and S. W. Cheong, Nature **429**, 392 (2004).

14) J. P. Rivera, Ferroelectrics **161**, 165 (1994).

15) Y. F. Popov, A. M. Kadomtseva, G. P. Vorob'ev, V. A. Sanina, A. K. Zvezdin, and M. M. Tehranchi, Physica B **284–288**, 1402 (2000).

16) Y. F. Popov, A. M. Kadomtseva, S. S. Krotov, D. V. Belov, G. P. Vorob'ev, P. N. Makhov and A. K. Zvezdin, Low Temp. Phys. **27**, 478 (2001).

17) A. K. Singh, D. Jain, V. Ganesan, A. K. Nigam and S. Patnaik, Euro. Phys. Lett. **86**, 57001 (2009).

18) T. Kimura, S. Kawamoto, I. Yamada, M. Azuma, M. Takano, and Y. Tokura, Phys. Rev. B **67**, 180401 (2003).

19) T. Katsufuji, S. Mori, M. Masaki, Y. Moritomo, N. Yamamoto, and H. Takagi, Phys. Rev. B **64**, 104419 (2001).

20) A. Singh, V. Pandey, R. K. Kotnala, and D. Pandey, Phys. Rev. Lett. **101**, 247602 (2008).

21) Z. J. Huang, Y. Cao, Y. Y. Sun, Y. Y. Xue and C. W. Chu, Phys. Rev. B **56**, 2623 (1997).

22) M. N. Iliev, H. G. Lee, V. N. Popov, M. V. Abrashev, A. Hamed, R. L. Meng and C. W. Chu, Phys. Rev B **56**, 2488 (1997).

23) M. A. Gilleo, Acta Cryst. **10**, 161 (1957).

24) H. L. Yakel, and W. C. Koehler, Acta Cryst. **16**, 957 (1963).

25) R. Feyerherm, E. Dudzik, N. Aliouane and D. N. Argyriou, Phys. Rev. B **73**, 180401 (2006).

26) M. Mostovoy, Phys. Rev. Lett. **96**, 067601 (2006).

27) M. Kenzelmann, A. B. Harris, S. Jonas, C. Broholm, J. Schefer, S. B. Kim, C. L. Zhang, S. W. Cheong, O. P. Vajk, and J. W. Lynn, Phys. Rev. Lett. **95**, 087206 (2005).

28) A. K. Singh, S. Patnaik, S. D. Kaushik and V. Siruguri, Phys. Rev. B **81**, 184406 (2010).

29) T. Lottermoser, T. Lonkai, U. Amann, D. Hohlwein, J. Ihringer, and M. Fiebig, Nature (London) **430**, 541 (2004).

30) J. W. Park, J. G. Park, G. S. Jeon, H. Y. Choi, C. H. Lee, W. Jo, R. Bewley, K. A. McEwen, and T. G. Perring, Phys. Rev. B **68**, 104426 (2003).

31) M. Fiebig, T. Lottermoser, D. Fröhlich, A. V. Goltsev, and R. V. Pisarev, Nature (London) **419**, 818 (2002).

32) S. Lee, A. Pirogov, J. H. Han, J. G. Park, A. Hoshikawa, and T. Kamiyama, Phys. Rev. B **71**, 180413 (2005).

33) S. Lee, A. Pirogov, M. Kang, K. H. Jang, M. Yonemura, T. Kamiyama, S. W. Cheong, F. Gozzo, N. Shin, H. Kimura, Y. Noda and J. G. Park, Nature **451**, 805 (2008).

34) P. A. Sharma, J. S. Ahn, N. Hur, S. Park, S. B. Kim, S. Lee, J. G. Park, S. Guha, and S. W. Cheong, Phys. Rev. Lett. 93, 177202 (2004).

35) A. Munoz, J. A. Alonso, M. J. Martinez Lope, M. T. Casais, and J. L. Martinez and M. T. Fernandez-Diaz, Phys. Rev. B **62**, 9498 (2000).

36) Y. Aikawa, T. Katsufuji, T. Arima, and K. Kato, Phys. Rev. B **71**, 184418 (2005).

37) S. Nandi, A. Kreyssig, J. Q. Yan, M. D. Vannette, J. C. Lang, L. Tan, J. W. Kim, R. Prozorov, T. A. Lograsso, R. J. McQueeney, and A. I. Goldman, Phys. Rev. B **78**, 075118 (2008).

38) T. Kimura, T. Goto, H. Shintani, K. Ishizaka, T. Arima and Y. Tokura, Nature **426**, 55 (2003).

39) B. Lorenz, A. P. Litvinchuk, M. M. Gospodinov, and C. W. Chu, Phys. Rev. Lett. **92**, 087204 (2004).

40) A. Muňoz, J. A. Alonso, M. J. Martinez-Lope, M. T. Casáis, J. L. Martinez and M. T. Fernández-Diaz, Chem. Mater. **13**, 1497 (2001).

41) S. Harikrishnan,S. Rößler, C. M. Naveen Kumar, H. L. Bhat, U. K. Rößler, S. Wirth, F. Steglich and Suja Elizabeth, J. Phys.: Condens. Matter **21**, 096002 (2009).

42) J. Park, M. Kang, J. Kim, S. Lee, K. H. Jang, A. Pirogov, and J. G. Park, Phys. Rev. B **79**, 064417 (2009).

43) R. V. Aguilar, A. B. Sushkov, C. L. Zhang, Y. J. Choi, S. W. Cheong, and H. D. Drew, Phys. Rev. B **76**, 060404 (2007).

44) M. Chandra Sekhar, S. Lee, G. Choi, C. Lee, and J. G. Park, Phys. Rev. B **72**, 014402 (2005).

45) T. Asaka, K. Nemoto, K. Kimoto, T. Arima, and Y. Matsui, Phys. Rev. B **71**, 014114 (2005).

46) Bas B. Van Aken, T. T. M. Palstra, A. Filippetti and N. A. Spaldin, Nat. Mater **3**, 164 (2004).

47) V. Y. Ivanov, A. A. Mukhin, A. S. Prokhorov, A. M. Balbashov, and L. D. Iskhakova, Phys. Solid State **48**, 1726 (2006);

48) N. Zhang, S. Dong, G. Q. Zhang, L. Lin,1 Y. Y. Guo, J.-M. Liu, and Z. F. Ren et al, Appl. Phys. Lett. 98, 012510 (2011).

49) Harikrishnan S. Nair, C. M. N. Kumar, H. L. Bhat, Suja Elizabeth, and Th. Br¨uckel, Phys. Rev. B. 83, 104424 (2011).

50) J. Rodriguez-Carvajal, Physica B **192**, 55 (1993).

51) N. A. Spaldin and M. Fiebig, Science **309**, 391 (2005).

52) T. Katsufuji, M. Masaki, A. Machida, M. Moritomo, K. Kato, E. Nishibori, M. Takata, M. Sakata, K. Ohoyama, K. Kitazawa, and H. Takagi, Phys. Rev B, **66** 134434 (2002).

53) G. Catalan Appl. Phys. Lett. 88, 102902 (2006).

**Figure Caption:**

**FIG 1.** (Color online) **(a)** X-ray diffraction pattern of polycrystalline $YMnO_3$ and $Y_{0.95}Dy_{0.05}MnO_3$ taken at room temperature. For clarity only small range of 2θ (28-34°) is shown. In the inset the whole 2θ range is shown for $Y_{0.95}Dy_{0.05}MnO_3$. All the peaks for both the samples are identified and indexed with hexagonal crystal structure and space group $P6_3cm$. All $YMnO_3$ peaks shift to the lower 2θ values due to doping of larger Dy in place of smaller Y. **(b)** ND pattern (symbols) taken at 300 K for $YMnO_3$ and $Y_{0.95}Dy_{0.05}MnO_3$. The solid line represents the calculated pattern with hexagonal symmetry. Inset shows the ND pattern taken at 10 K for both the samples. The magnetic Bragg peaks are marked by (*). The lines below the bars in the figure indicate the difference between the observed and calculated diffraction patterns. The bars indicate the position of the nuclear Bragg peaks. The upper and lower sets of bars in the inset correspond to nuclear and magnetic phases, respectively.

**FIG 2.** (Color online) Temperature dependence of **(a)** lattice constant ***a***, lattice constant ***c*** **(b)** unit cell volume **(c)** ordered magnetic moment of Mn atom

**FIG 3.** (Color online) (a)-(d**)** Mn-O bond lengths, **(e)-(j)** O-Mn-O bond angles. All the O-Mn-O bond angles are measured in degrees. All the values are obtained from Rietveld refinement of ND pattern at various temperature for pure $YMnO_3$ (■) and $Y_{0.95}Dy_{0.05}MnO_3$ (Δ). Here O1, O2 represent apical oxygen atoms and O3, O4 represent equatorial oxygen atoms.

**FIG 4.** (Color online) Temperature variation of the magnetic susceptibility measured in zero field-cooled (ZFC) conditions under $\mu_0H = 0.5$ T is shown for $YMnO_3$ (■) and $Y_{0.95}Dy_{0.05}MnO_3$ (•). Inset (a) and (b) show the temperature dependence of inverse susceptibility for $YMnO_3$ and $Y_{0.95}Dy_{0.05}MnO_3$ respectively. Symbols are for data point and the line is a fitting result using the Curie-Weiss law.

**FIG 5.** (Color online) **(a)** Temperature dependence of dielectric constant measured at $\mu_0H$ = 0 T for $YMnO_3$ (■), and $Y_{0.95}Dy_{0.05}MnO_3$ (●) taken at 1 kHz frequency in the temperature range 5-250 K. Inset shows the inverse S-shape anomaly for $YMnO_3$ (at ~ 62 K) and $Y_{0.95}Dy_{0.05}MnO_3$ (at ~ 58 K) which is a signature of AF transition. **(b)** Temperature dependence of dielectric constant measured at different magnetic field $\mu_0H$ = 0 T (▬), 3 T (▬) and 5 T (▬) taken at 1 kHz frequency for $YMnO_3$ (upper panel) $Y_{0.95}Dy_{0.05}MnO_3$ (lower panel) are shown.

**FIG 6.** (Color online) **(a)** The temperature dependence of specific heat (raw data) for $YMnO_3$ (upper panel) and $Y_{0.95}Dy_{0.05}MnO_3$ (lower panel) at $\mu_0H$= 0 T are shown. The anomaly in C/T plot for both the samples is a clear signature of AF ordering. The inset of each panel shows the magnetic contribution of specific heat which is obtained from raw data of specific heat after subtraction of phonon contributions. **(b)** Temperature dependence of specific heat for $Y_{0.95}Dy_{0.05}MnO_3$ done at $\mu_0H$= 0 T (▬), 1 T (▬) and 5 T (▬) are shown. Inset shows the clear variation in the heat capacity at low temperatures.

## Table caption

**Table 1.** Structural parameters after the Rietveld refinement of neutron diffraction pattern of $YMnO_3$ and $Y_{0.95}Dy_{0.05}MnO_3$ at room temperature. Atomic positions of Y1 (Dy1) and O3 at 2a (0,0,z); Y2 (Dy2) and O4 at 4b (1/3,2/3,z); Mn, O1, and O2 at 6c (x,0,z) (for Mn, z = 0).

**Table 1**

| Parameters | | $YMnO_3$ | $Y_{0.95}Dy_{0.05}MnO_3$ |
|---|---|---|---|
| a (Å) | | 6.1490 | 6.1686 |
| c (Å) | | 11.3627 | 11.3345 |
| V ($Å^3$) | | 372.073 | 373.511 |
| Y1 | Z | 0.27206 | 0.27206 |
| | Biso | 0.422 | 0.422 |
| Y2 | Z | 0.22900 | 0.22900 |
| | Biso | 0.422 | 0.422 |
| Mn | X | 0.32472 | 0.32472 |
| | Biso | 0.498 | 0.498 |
| O1 | X | 0.30952 | 0.30952 |
| | Biso | 0.450 | 0.450 |
| | Z | 0.15127 | 0.15127 |
| | Biso | 0.450 | 0.450 |
| O2 | X | 0.64491 | 0.64491 |
| | Biso | 0.445 | 0.445 |
| | Z | 0.32599 | 0.32599 |
| | Biso | 0.445 | 0.445 |
| O3 | Z | 0.47814 | 0.47814 |
| | Biso | 0.795 | 0.795 |
| O4 | Z | 0.01227 | 0.01227 |
| | Biso | 0.795 | 0.795 |
| Discrepancy factors | | | |
| $\chi^2$ | | 3.93 | 3.82 |
| $R_p$ (%) | | 2.17 | 3.10 |
| $R_{wp}$ (%) | | 2.89 | 4.05 |
| $R_{exp}$(%) | | 1.23 | 2.09 |

**Fig 1(a)**

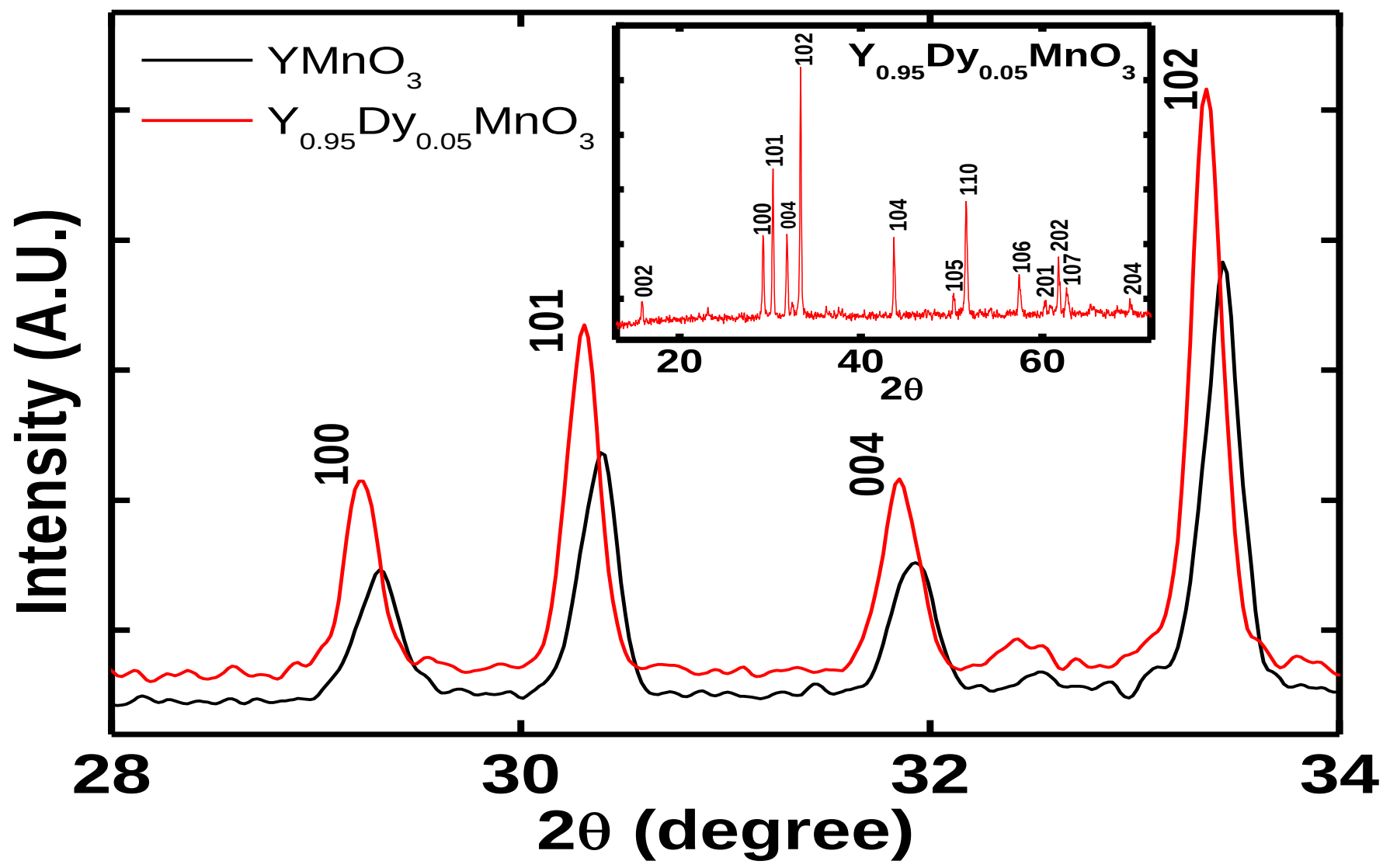

**Fig 1(b)**

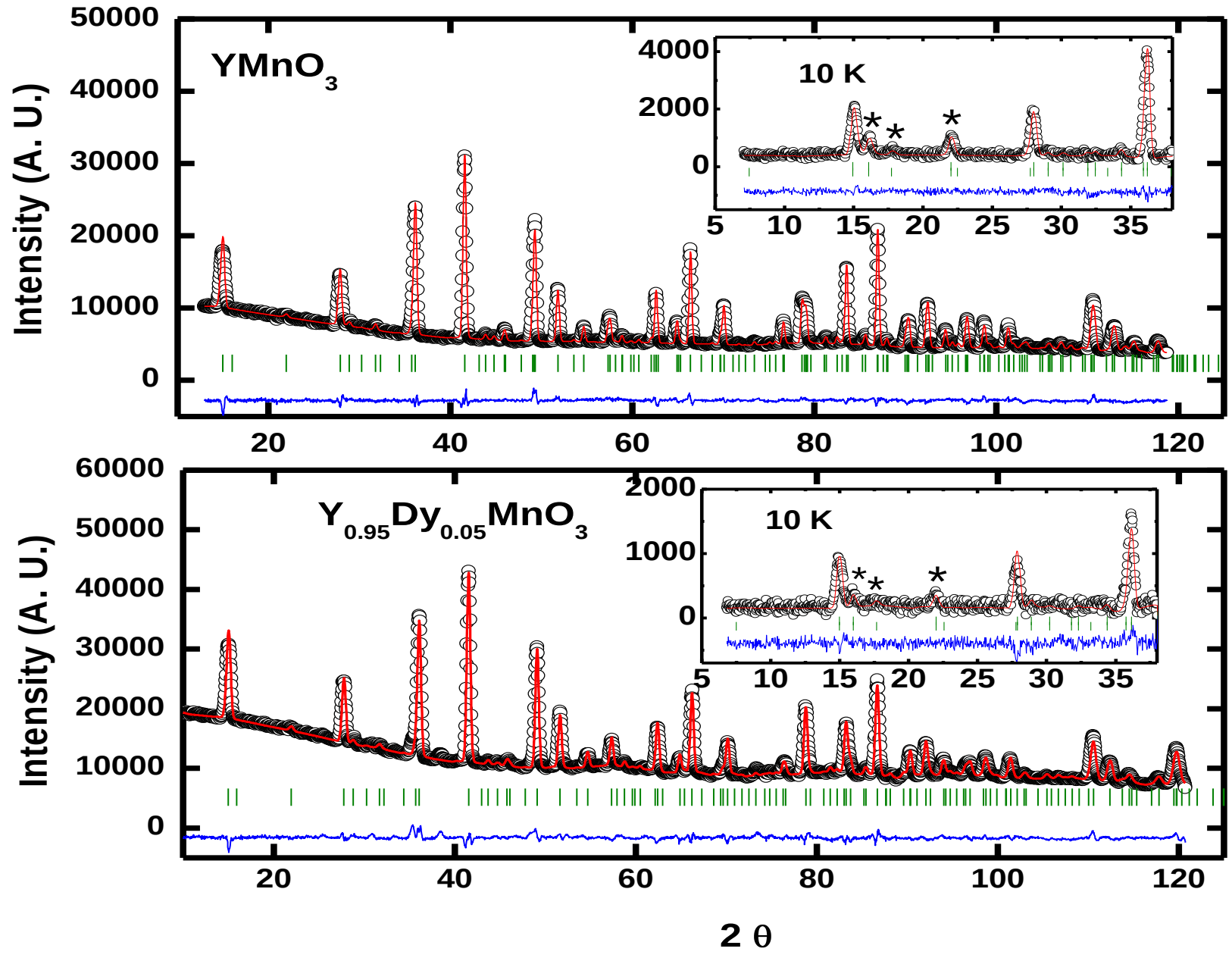
YMnO3
Y0.95Dy0.05MnO3
10 K
Intensity (A. U.)
2 θ

**Fig 2.**

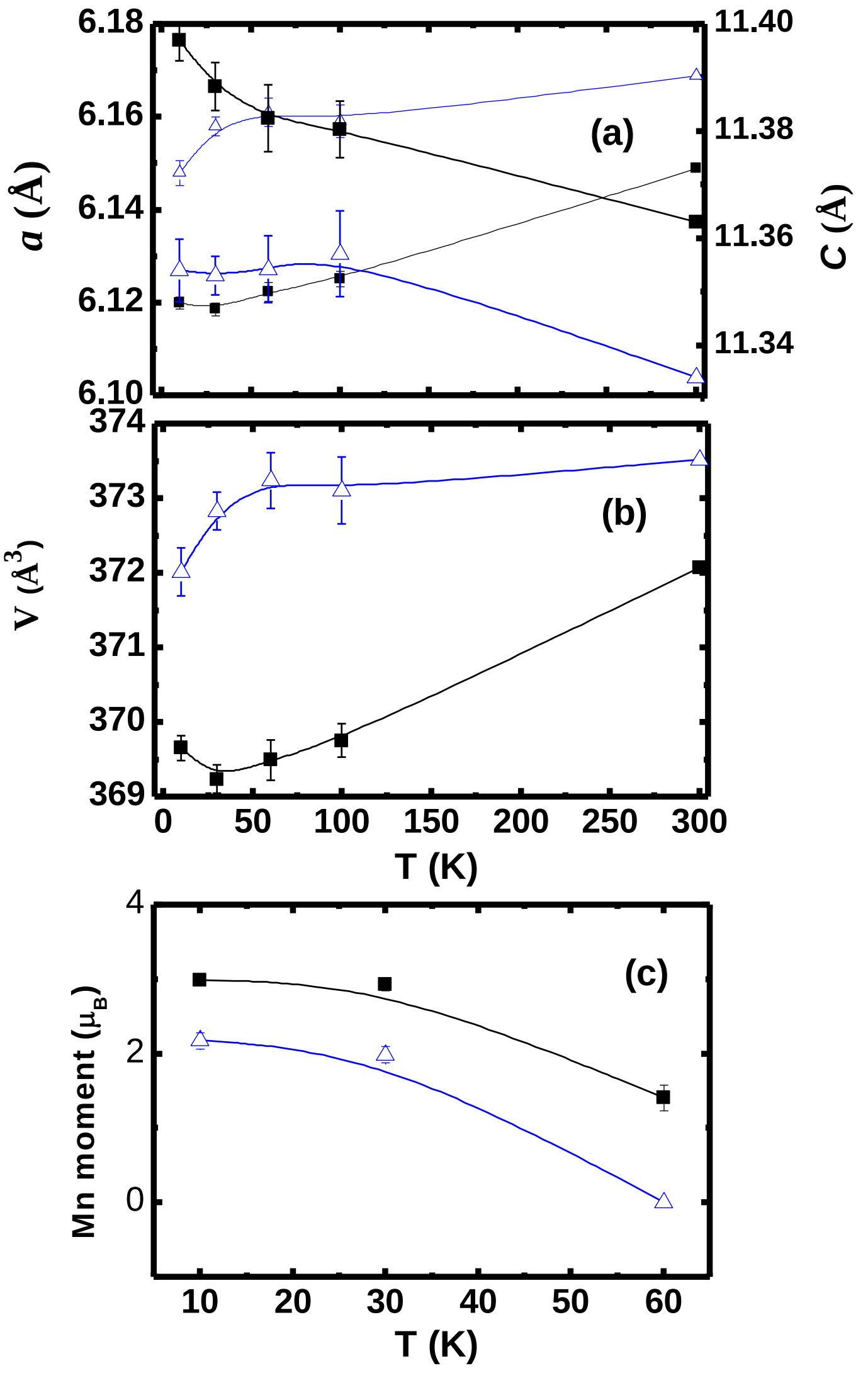

**Fig 3.**

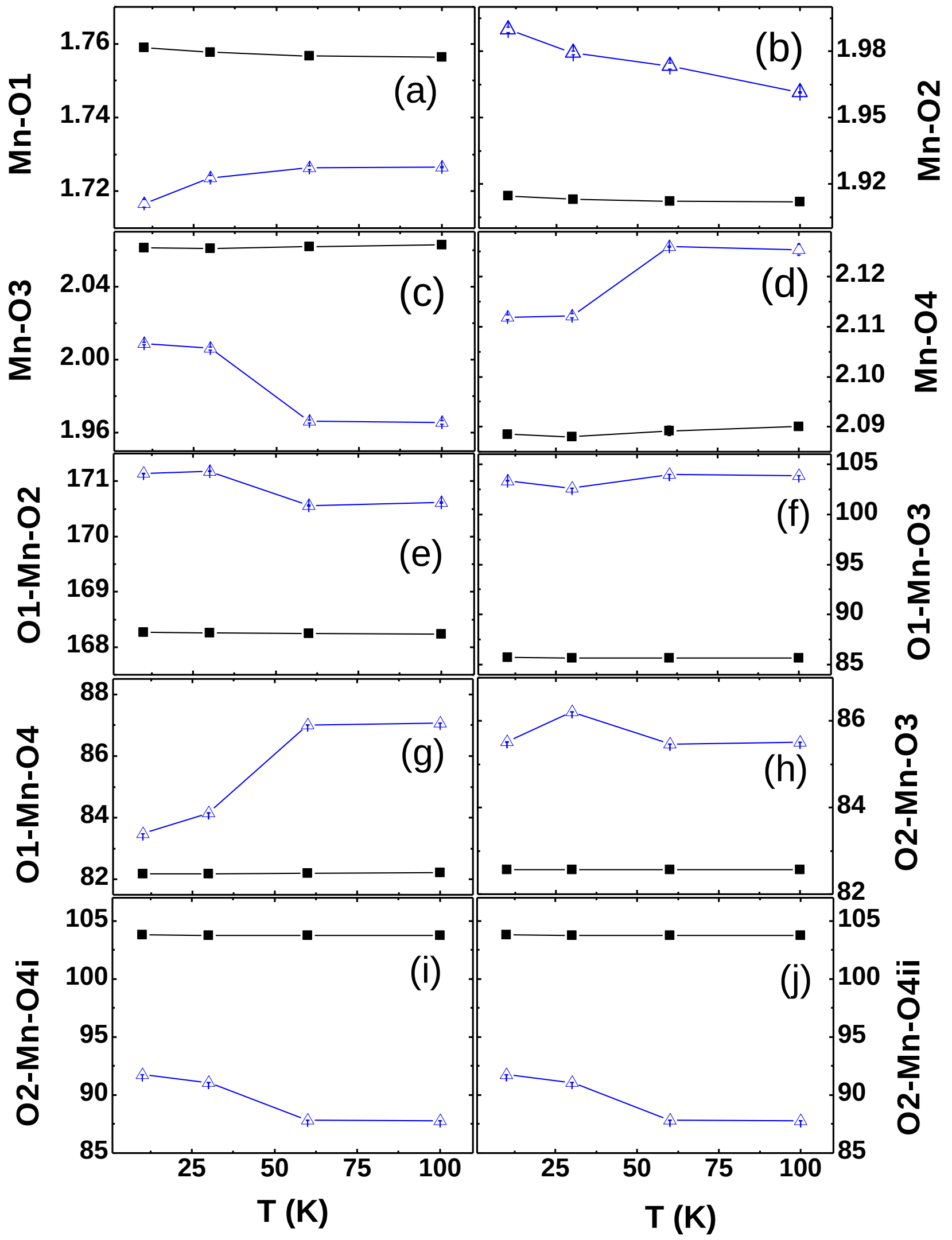

**Fig 4.**

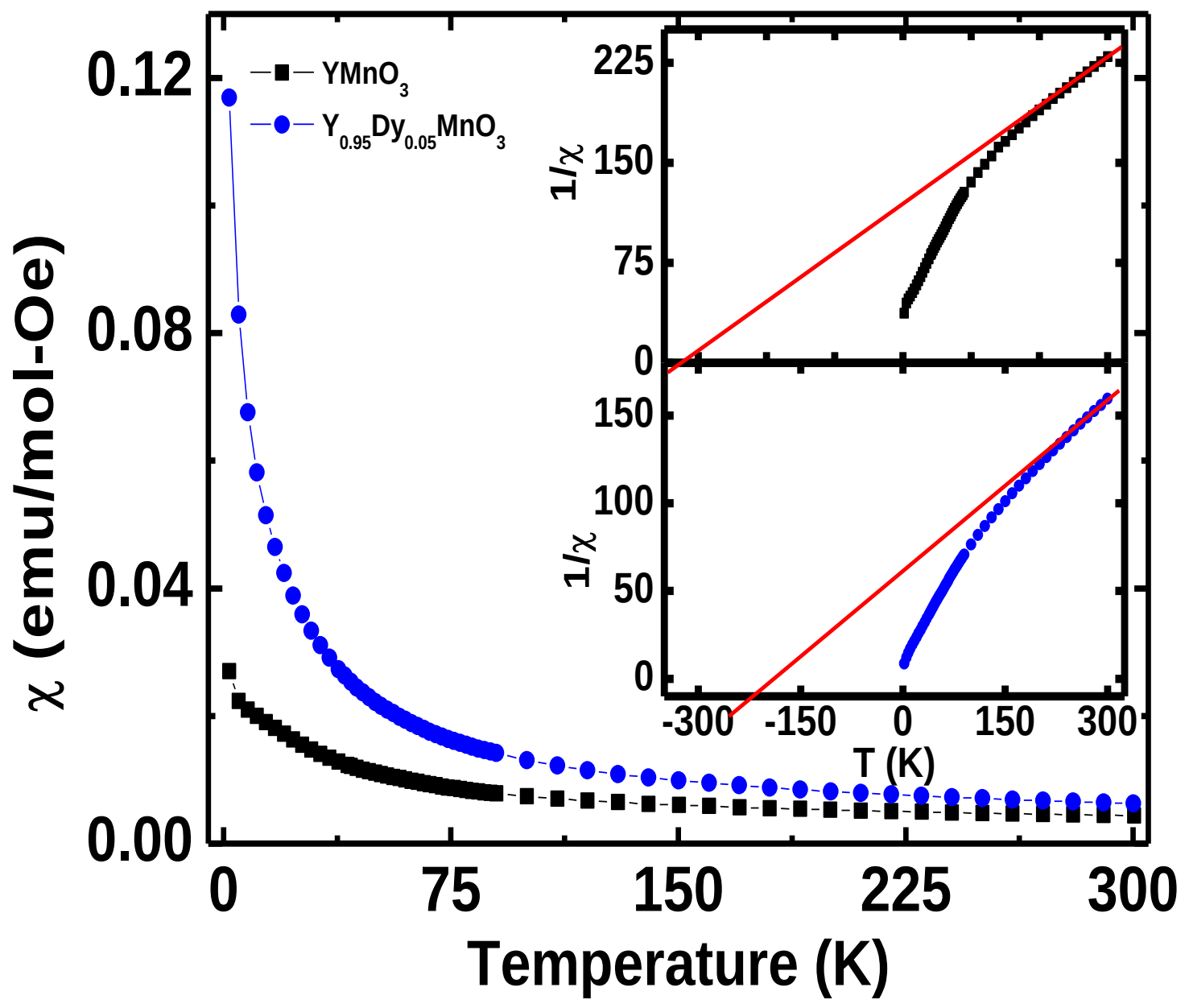

Fig 5(a)

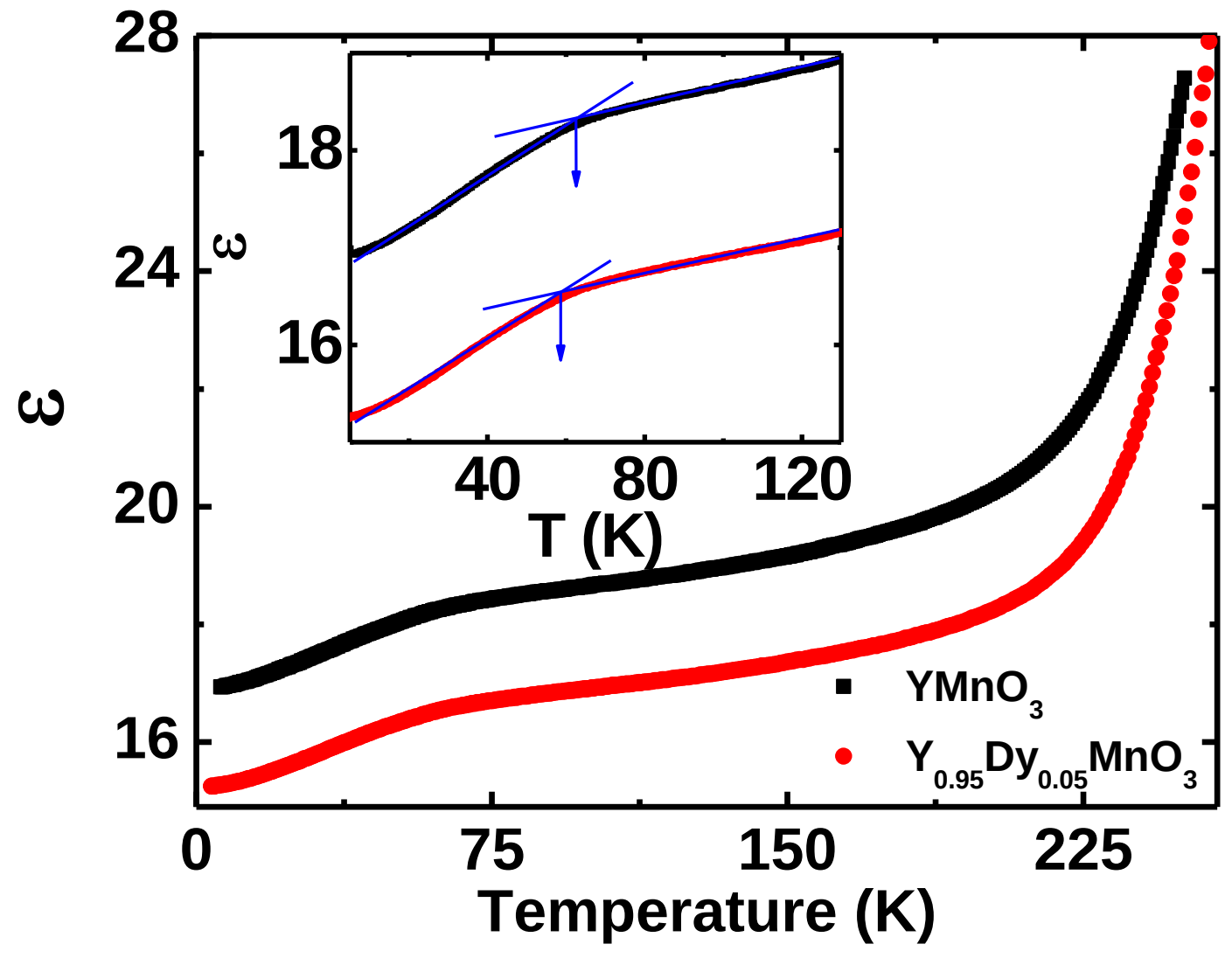

Fig 5(b)

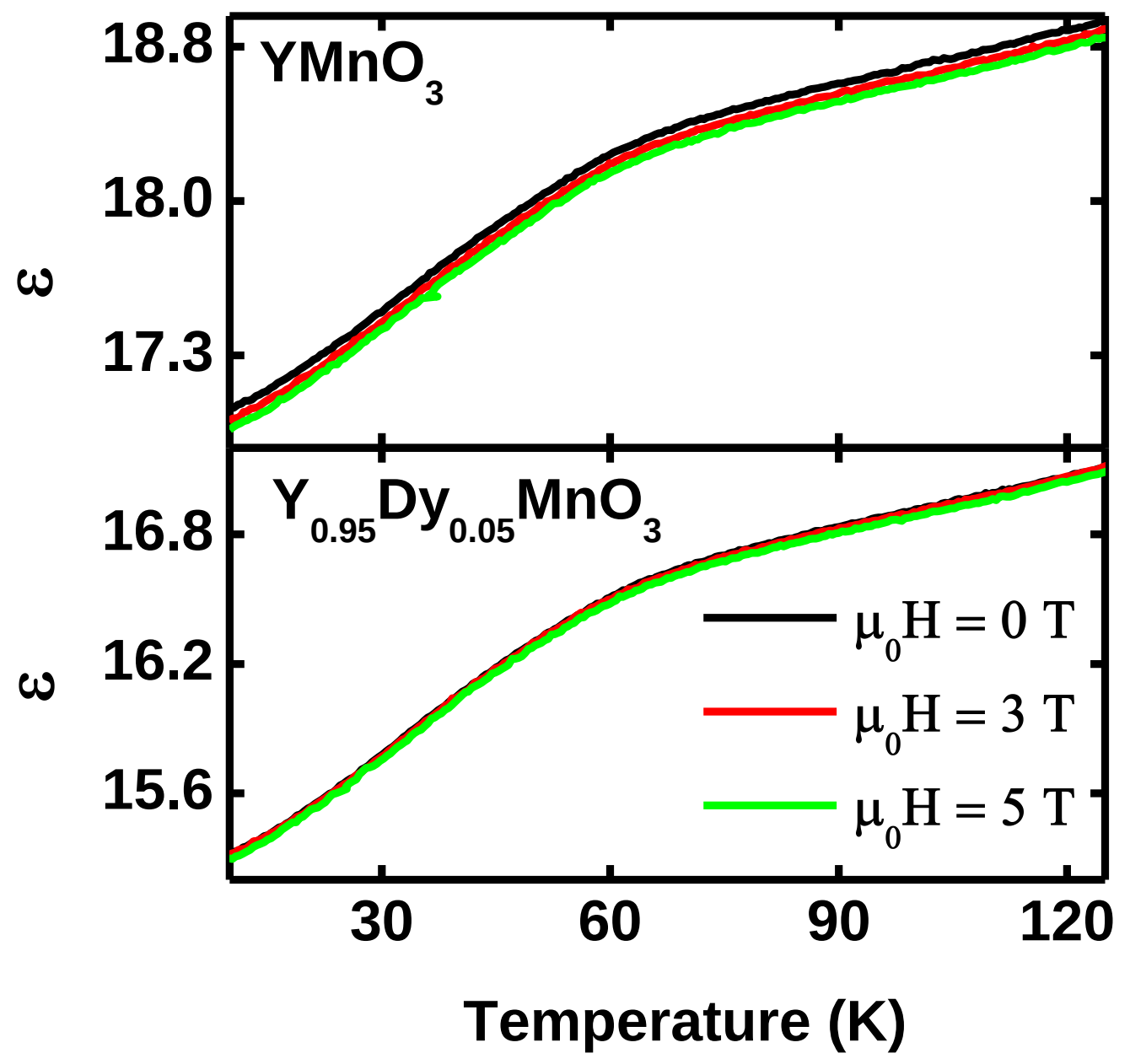

Fig 6(a)

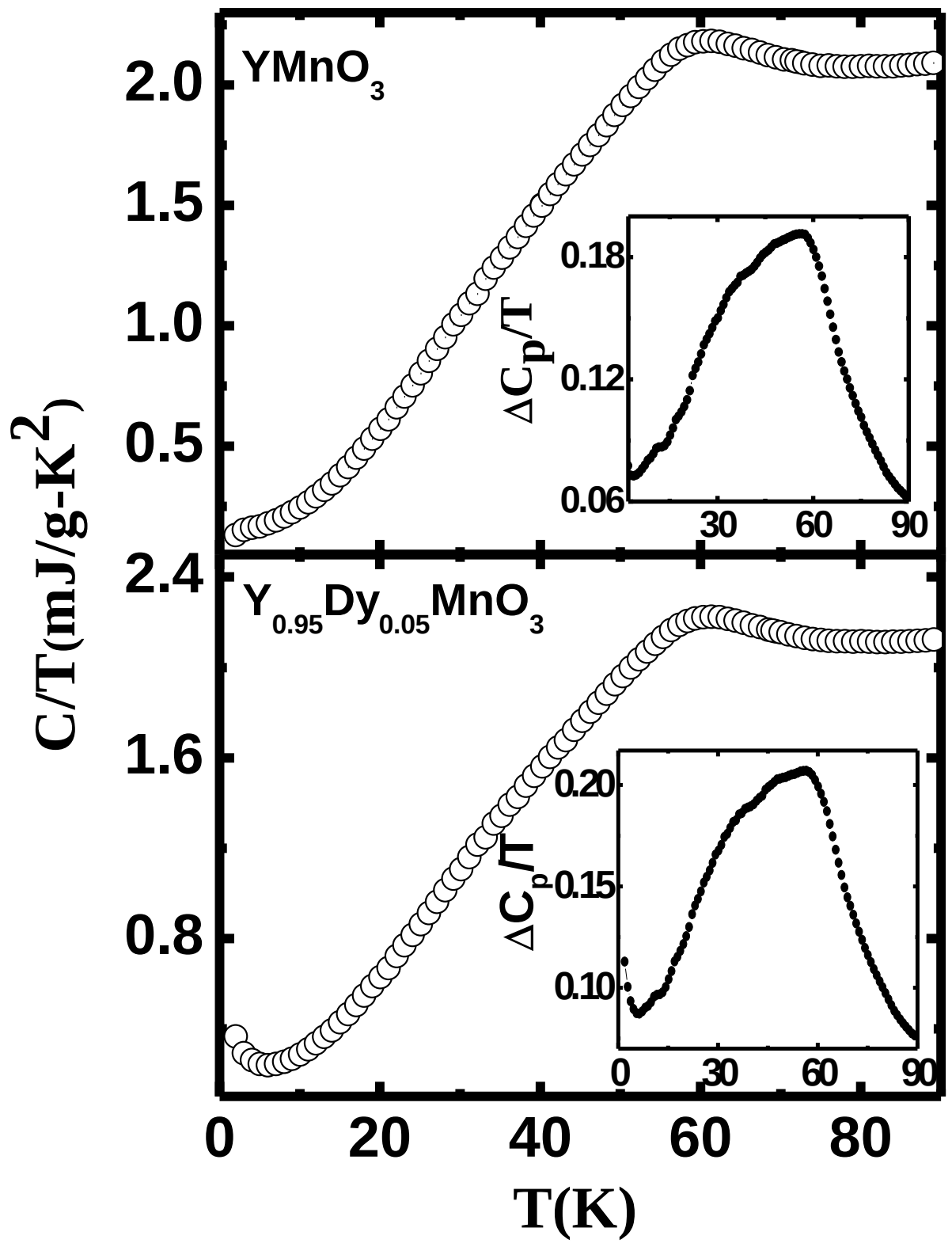

Fig 6(b)

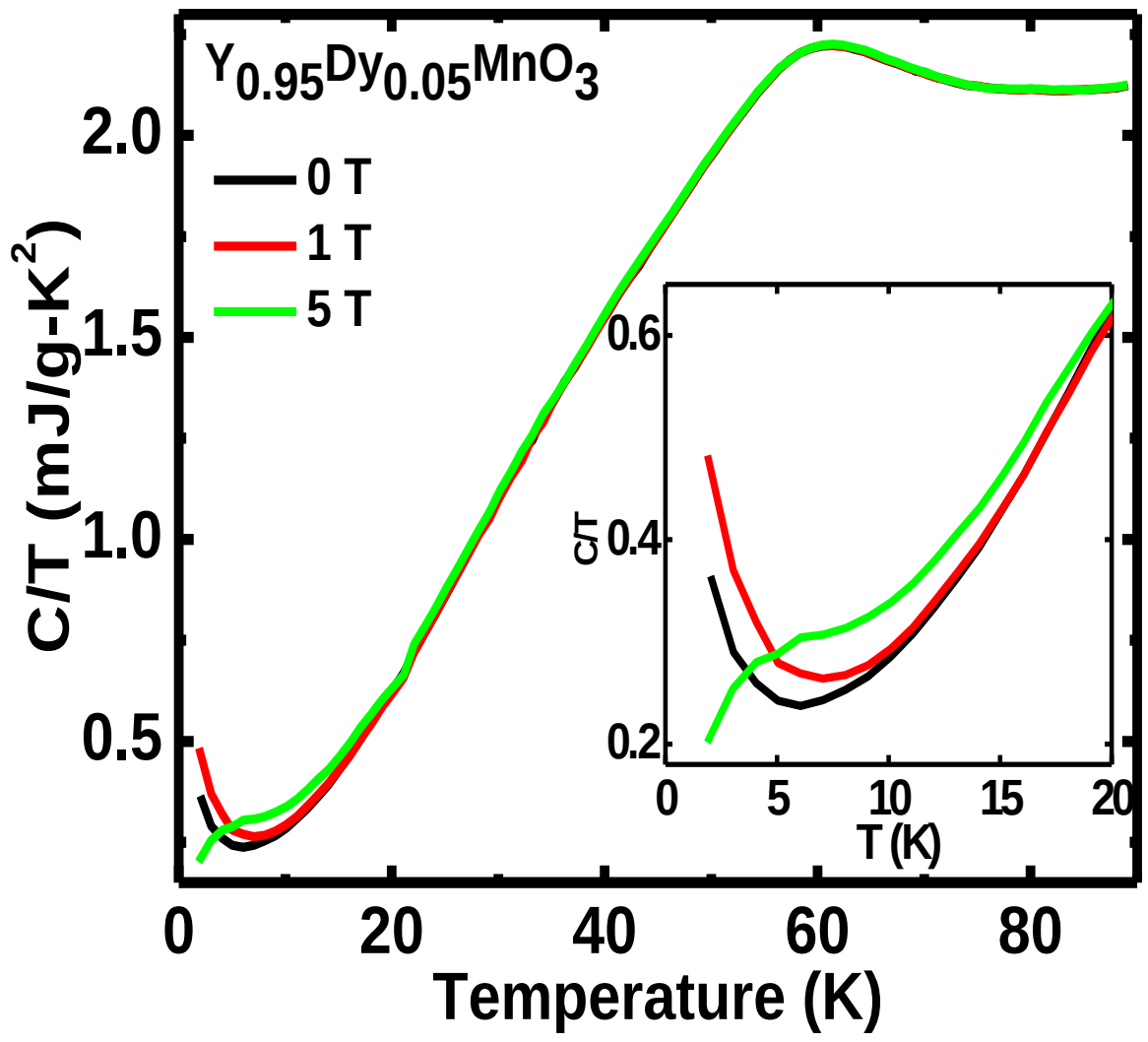